\documentstyle [preprint,aps]{revtex}

\newcommand {\vol} [1] {{\bf #1}}
\begin{document}

\title{Influence of electron phonon interaction on superexchange}
\author{W. Stephan,  M. Capone, M. Grilli, and C. Castellani}
\address{Dipartimento di Fisica, Universit\'a La Sapienza,
Piazzale Aldo Moro, 00185 Roma, Italy}
\maketitle

\begin{abstract}
We investigate the influence of electron-phonon coupling
on the superexchange interaction of magnetic insulators.  
Both the Holstein-Hubbard
model where the phonons couple to the electron density, as well
as an extended Su, Schrieffer, Heeger model where the coupling arises
from  modulation of the overlap integral are studied using exact
diagonalization and perturbative methods.  In all cases for
both the adiabatic (but non-zero frequency) 
and anti-adiabatic parameter regions the
electron-phonon coupling is found to enhance the superexchange.
\end{abstract}

\vfill \eject
There is by now a large body of work which demonstrates that the
electronic structure of the high temperature superconducting cuprates 
may be quite well described by a tight-binding model with intermediate
to strong Coulomb interactions.  
On the other hand, there are also experimental results which
support the presence of sizeable electron-phonon
interactions, in particular for low hole doping\cite{CC}.
One may then ask how strongly the inclusion of electron-phonon interactions
affects the models in question.  In particular, the value of
the superexchange interaction $J$ is rather directly accessible
experimentally, and constrains the parameters which may be used
in any model description.  If the effective superexchange constant
has a significant dependence on the electron-phonon interaction,
this could have significant consequences for the models used.

We have approached these questions by examining two distinct models
for the electron-phonon coupling:  a) the Holstein-Hubbard
\cite{Holstein,polaron_review}
 (HH) model,
where Einstein phonon modes are assumed to be coupled to the
local density (fluctuations) of the electrons, and b) the extended
Su, Schrieffer, Heeger (SSH) model\cite{SSH}, where the ionic displacements are
assumed to modulate the electronic overlap integrals.  
In the present work we
restrict ourselves to only a single band model for the
electronic degrees of freedom.

%
This question has previously been considered in Ref.
\cite{ZS,proc}
for the Holstein-Hubbard model in the extreme adiabatic limit
$\omega_0 \rightarrow 0$, where it was found that
there is no dependence of the superexchange on the electron-phonon
coupling.  This occurs because there is no
time for any lattice motion to take place during the virtual
hopping process which lowers the singlet state energy relative to
that of the triplet.   
On the other hand, Ihle and Fehske\cite{IF} have considered also the
non-adiabatic limit within a variational approach and have found
some non-trivial dependence of $J$ on the electron-phonon coupling.
According to the standard approach to
the small polaron problem, the Lang-Firsov\cite{LF} canonical transformation,
the effective hopping integral is exponentially suppressed
at strong coupling. Naively this would lead to a large reduction
of $J$. At the same time however, the effective
on-site Coulomb repulsion is reduced due to the attractive
interaction which eventually leads to on-site bipolaron
formation at sufficiently strong coupling. 
This leads to a reduced energy denominator for the intermediate
states, which would tend to enhance the superexchange.
Thus from this point of view one would expect two  different effects
which go in opposite directions, and {\it{a priori}} the final result
is not obvious.

In this paper
the effective exchange interaction is determined using both 
exact diagonalization
of small clusters, truncating the Hilbert space by allowing
only a finite maximum number of phonons per mode\cite{RT,AKR}, 
as well
as by approximate methods combining canonical transformations
and/or perturbation theory in the spirit of the above discussion.
Both models display a common trend:  $J$
is enhanced with increasing electron-phonon coupling.
Indeed the reduction of the hopping integral well known for the
single polaron problem does not manifest itself in the determination of $J$.
The magnitude of the enhancement is generally quite small
in the Holstein-Hubbard model, case a), but can be more significant in
case b).  The fact that the influence of the electron-phonon
interaction is greater in the SSH model than in the HH model
is symptomatic of a
clear difference in the character of spin-phonon coupling in
the two models which will be clarified further in this paper.

%
%

The HH model is defined by
$H = H_E + H_P + H_{EP}$ where
\begin{equation}
H_E + H_P = -t\sum_{<i,j>\sigma}(c^\dagger_{i\sigma} c_{j\sigma} + 
                       c^\dagger_{j\sigma} c_{i\sigma} )
 +  U\sum_i n_{i\downarrow}n_{i\uparrow}
 + \omega_0 \sum_i a^\dagger_i a_i
\label{heandp}
\end{equation}
and
\begin{equation}
H_{EP} = g \sum_i (n_i-<n>)(a^\dagger_i + a_i).
\label{hep}
\end{equation}
Here $c^\dagger_{i\sigma} (c_{i\sigma})$ are creation (annihilation)
operators for electrons of spin $\sigma$ and $a^\dagger_i (a_i)$
are creation (annihilation) operators for local phonons.
In (\ref{heandp}) $<i,j>$ implies a sum over nearest neighbor pairs
with $j > i$.
This model describes tight-binding electrons interacting
with a local Coulomb repulsion $U$, coupled to Einstein oscillators
of frequency $\omega_0$ via the local electron density.
For the present purpose it is useful to consider coupling to
the fluctuations of electronic density $n_i - <n>$, where
$<n>$ is the mean electron density, rather than to the electronic density
$n_i$ directly.  This eliminates the trivial coupling of
the zero momentum phonon mode to the total electron density\cite{RT},
which reduces the number of phonons required to achieve convergence.
The zero-point energy of the phonons is constant and will be suppressed,
and throughout we use units such that $\hbar = 1$.
For the SSH model the electron-phonon interaction (\ref{hep}) is
replaced by a modulation of the electronic overlap integral due
to the lattice displacement, described by
\begin{equation}
H_{EP^\prime} = g \sum_{<i,j>\sigma}\left(c^\dagger_{i\sigma}c_{j\sigma} +
 h.c.\right)
  \left(a^\dagger_j + a_j - a^\dagger_i - a_i \right).
\label{Hg}
\end{equation}

The most straightforward way to estimate the superexchange
interaction $J$ is to solve ``exactly''
for the lowest singlet ($E_0$) and triplet ($E_1$) eigenvalues for two
electrons on a two site cluster.  
For sufficiently weak electron-phonon coupling and large $U$
such that electronic configurations
with only a single electron per site predominate, the difference
$E_1 - E_0$ may be interpreted as the effective
exchange interaction $J$.  Of course for larger coupling $g$
one observes a transition to a local bipolaron, in which case
this energy difference has a different physical interpretation,
ie. the bipolaron binding energy.
Note that with our choice of coupling the phonons
to the density fluctuations the triplet state does not couple
to the phonons, and hence we always have $E_1 = 0$.
Therefore to determine the exchange interaction we only need
the singlet ground state energy $E_0$, which will be obtained
by direct diagonalization using the Lanczos algorithm\cite{AKR} 
as well as within several approximation schemes.
It is at this point that the analysis for the HH
model is simpler than for the SSH model:  in principle in both cases there
are also corrections to this estimate of $J$ due to processes
involving more than two sites.  From the low-lying energies of a three site
cluster one may readily extract the nearest-neighbor ($J$) and 
next nearest neighbor ($J^\prime$) exchange couplings of an effective
spin model\cite{JJWS}.  
The difference between $J$ determined from three sites and that extracted
from only two sites may be shown to be the leading correction in a
systematic expansion.  We have calculated these leading 
corrections to the effective $J$ found from two sites 
and find that in the large $U$ limit they
are negligible for the HH model, even for rather large $\omega_0$
and $g$, whereas the non-local coupling of the SSH model
leads to very significant corrections for strong electron-phonon
coupling.  The physical mechanisms leading to this difference 
will be discussed further in simple terms using strong-coupling
perturbation theory.

Fig. 1(a) summarizes the dependence of $J$ on the electron-phonon
coupling $g$ in the HH model for a fixed value of $U$ and a range of phonon
frequencies.  The dotted lines diverging upwards indicate the continuation
into the region of an on-site bipolaron, where of course our
identification of $J$ with the singlet-triplet splitting
is no longer physically meaningful.  One can note here that 
the onset of bipolaron formation  agrees very well with the
prediction of the Lang-Firsov approach for this instability at
$\tilde{U} = U - 2 g^2/\omega_0 \rightarrow 0$ for all
phonon frequencies.
In particular it is
worth noting that this result, which is exact in the extreme
antiadiabatic limit ($t/\omega_0 \to 0$), also holds in the
adiabatic limit ($\omega_0 /t \to 0$) \cite{ZS}.
   The solid lines show that the superexchange
is enhanced by increasing electron-phonon coupling, with
the maximum enhancement increasing with the phonon frequency.

In order to understand better these numerical results we now
derive simple approximate forms for the ground state
energy of the two site cluster.
We first eliminate coupling of phonons to electron density
 using the standard Lang-Firsov canonical
transformation\cite{LF}, with a small generalization to accomodate our form
of electron-phonon coupling.  The generator of the transformation is
$S = -\alpha \sum_i(n_i-<n>)(a_i - a^\dagger_i)$
which transforms (\ref{heandp}) and (\ref{hep}) to $H = H_0 + H_t$, where
\begin{equation}
H_0 = -\alpha^2\omega_0\sum_i\left( n_i - <n> \right)^2 + 
      \omega_0\sum_i a^\dagger_i a_i
      + U\sum_i n_{i\downarrow}n_{i\uparrow}
\end{equation}
is diagonalized by the transformation, and the kinetic energy term
becomes
\begin{equation}
H_t = -t\sum_{<i,j>\sigma}\left( c^\dagger_{i\sigma}c_{j\sigma}X^\dagger_i
  X_j + h.c. \right).
\label{ht}
\end{equation}
Here $X_i = exp\left[ \alpha(a_i - a^\dagger_i) \right]$
and $\alpha = g/\omega_0$.
At this point, in order to proceed further one has at least two
alternatives:  1) the hopping term (\ref{ht}) can be treated as a perturbation
and standard Rayleigh-Schr\"odinger perturbation theory to second
order may be performed, or 2) a further canonical transformation of
the Schrieffer-Wolf type\cite{Harris_Lang} 
may be used to eliminate to order $t/U$
processes which change the number of doubly occupied sites.
In the absence of phonons these two procedures are completely equivalent,
however in the present case they differ due to terms involving multi-phonon
processes which are not included in the Schrieffer-Wolf transformation
while they are taken into account in Rayleigh-Scr\"odinger perturbation
theory.
Even if $U \gg \omega_0$ the difference
between the two can be significant for strong electron-phonon coupling.

In the case 2), to lowest order in the hopping (\ref{ht}) one simply
obtains $J_{SW} = 4t^2/\tilde{U}$, i.e. the effective Coulomb repulsion
is reduced from the bare value
by the bipolaronic attraction.  This result agrees with the one obtained
by Ihle and Fehske\cite{IF} along the same line.  
We shall see from our exact results that this is a poor approximation.
On the other hand, performing 
second order perturbation theory in the hopping (\ref{ht})
leads to
\begin{equation}
\label{jpt}
J_{PT} = {4t^2\over U} \exp(-2\alpha^2)\sum_{n,m=0}^\infty \left[
\alpha^{2(n+m)} \over
{n! m! (1 - {2\alpha^2\omega_0\over U}  + 
{{(n + m)\omega_0} \over U}})\right]
\end{equation}
Although (\ref{jpt}) is only second order perturbation theory in $t/U$
it includes some effects of e-ph couplig to all orders.
The sum in (\ref{jpt}) is readily evaluated numerically.   Note that in the
limit $\omega_0 \rightarrow 0$ with $g^2/\omega_0$ remaining finite
we recover the
adiabatic result $J=4t^2/U$.
This adiabatic result may also be found by first performing 
the Schrieffer-Wolf transformation to $O(t/U)$ followed by
the Lang-Firsov transformation.  Within this approach
to generate a dependence of $J$ on electron-phonon coupling when $\omega_0$
is not zero (as shown by Fig. 1 and equation (\ref{jpt})
it is necessary to go beyond leading
order in the Schrieffer-Wolf transformation.

Fig. 1(b) shows a comparison of the exact diagonalization results
with $J_{SW}$ and $J_{PT}$ for one choice of phonon frequency
and a pair of different values for the Coulomb repulsion.
The agreement between $J_{PT}$ and
the exact results is quite good, whereas $J_{SW}$ tends to overestimate
the enhancement.  
Note also that the adiabatic approximation result
(J independent of g) is better than $J_{SW}$ even for $\omega_0$ of order $t$.
This finding is in agreement with the results of Ref.\cite{proc}
where it was argued that the lattice has no time to relax for
$\omega_0 < U,U_b=2\alpha^2\omega_0$, making an adiabatic
treatment appropriate.
It is therefore not surprising that the adiabatic result holds
for $\omega_0 \sim t$.
One may summarize these
results by saying that the superexchange is only slightly
enhanced due to strong electron-phonon coupling in the 
HH model, until very near the coupling value
where the Coulomb repulsion is overcome by the effective
attractive interaction and an on-site bipolaron is formed.
This result contrasts with reference \cite{IF} according to which
a decrease of $J$ is possible for some parameters.
If one attempts to reproduce the exact results using simple
canonical transformations or perturbation theory, the optimal
scheme appears to be to perform a Lang-Firsov transformation,
followed by ``old fashioned'' second order perturbation theory
in the transformed hopping.  This simple procedure leads to a
better treatment of the $\omega_0/U$ terms than is the case
when a Schrieffer-Wolf transformation to leading order
{\it after the Lang-Firsov transformation} 
is performed instead of the perturbation calculation.

Next let us consider the SSH model:
Fig. 2(a) shows the dependence of the effective exchange interaction
on the electron phonon coupling for  the SSH model.
As in Fig. 1(a), the dotted lines indicate the extension of the
curves into the bipolaron phase.
Note that in this model a single polaron in the strong-coupling limit
is quite well described as an electron localized on a shortened
bond.  The bipolaron here is then also not localized on a single
site as in the HH model, but rather on a bond.  Therefore in this
case the distinction between the ``singly-occupied'' spin-model
and the bipolaron phase is not as clear cut.  Nevertheless, the
expectation value of double occupancy shows a significant change
at a rather well defined coupling that we take to define the
beginning of the bipolaron phase.
Note that the size of the enhancement of $J$ seen here can be larger than
that of the HH model in Fig. 1(a).
For this model we are not aware of any canonical transformation analagous
to the Lang-Firsov of the Holstein model.  In order to derive simple
approximations for the singlet ground state energy we therefore
compare only the Schrieffer-Wolf approach and direct perturbation 
theory in the hopping and electron phonon interaction.  
The latter is quite trivial to second order, and leads to the
prediction $J = 4[t^2/U + 2g^2/(U+\omega_0)]$.  On the other hand,
performing the Schrieffer-Wolf transformation to leading order leads
at half-filling to a Heisenberg model with spin-phonon coupling
\begin{eqnarray}
\tilde{H} & = & 4t^2/U \sum_{<i,j>}({\vec S}_i\cdot{\vec S}_j - 1/4)
  -8tg/U \sum_{<i,j>}({\vec S}_i\cdot{\vec S}_j - 1/4)
    (a^\dagger_j+a_j- a^\dagger_i- a_i) \nonumber\\
   & + & 4g^2/U \sum_{<i,j>}({\vec S}_i\cdot{\vec S}_j - 1/4)
    (a^\dagger_j+a_j- a^\dagger_i- a_i)^2
 + \omega_0 \sum_i a^\dagger_i a_i.
\label{hspin}
\end{eqnarray}
Note that this amounts to considering the purely electronic $J = 4t^2/U$
and taking $t \rightarrow t-g(a^\dagger_j+a_j- a^\dagger_i- a_i)$.
The two electron singlet ground state of (\ref{hspin}) for two sites
may be solved exactly, leading to 
\begin{equation}
J = 4t^2/U + 8g^2 e^{-2\gamma}/U -\omega_0 \sinh^2\gamma
    - 128(tg)^2/(32g^2U - \omega_0 U^2)
\label{JSSH}
\end{equation}

where $\gamma$ is the solution of $\coth2\gamma = 1 - \omega_0 U/(16g^2)$.
This result is compared to the numerical result of the 
model (\ref{Hg}) in fig. 2(b).
Within the effective spin model the phonon mode is predicted to
go soft at $g^2 = \omega_0 U/32$.  From the direct numerical 
diagonalization of the original model (\ref{Hg}) this coupling value
is found to be in quite good agreement with the onset of the
formation of a short-range bipolaronic state as determined by the
behavior of the double occupancy discussed above.
At this point the difference between the SSH model and the HH model
as regards spin-phonon coupling is evident:  recall that in the
HH model the leading order SW transformation leads to no explicit
spin-phonon coupling.  This difference is perhaps most simply
understood from the point of view of strong-coupling perturbation
theory where $U \gg (t$, $g)$.  Consider first the limit $g \rightarrow 0$:
the superexchange $4t^2/U$ arises due to one application of the
hopping term, leading to intermediate states with one empty and
one doubly occupied site of energy $U$, followed by a further
hop leading to a configuration with only single occupancy.
In the HH model, the leading correction to this well known
result arising from the electron-phonon coupling involves
first a hop creating a doubly occupied site with energy $U$, but now followed
by the application of $H_{EP}$ creating a phonon, leading to
a state of energy  $U+\omega_0 \simeq U$.  The phonon can then be
annihilated and the double occupancy removed by the reverse
process, giving a leading order correction $O(t^2g^2/(U^3)$, in agreement 
with (\ref{jpt}).
Similar analysis of higher order processes leads to the conclusion
that in terms of powers of $t/U$ or $g/U$ the electron-phonon corrections
always begin at higher order than the comparable terms of the Hubbard
model.  In the case of the SSH model however, the leading
correction to $J$ arises already at order $g^2/U$: application
of (\ref{Hg}) leads to an intermediate state of energy $U+\omega_0$,
and a second application of (\ref{Hg}) returns to the ground manifold.
This term is one of those appearing in (\ref{hspin}).
Quite generally, corrections due to the electron phonon coupling
begin at the same order in $1/U$ as similar terms due to the
hopping, leading to large corrections if $g/t$ is not small.
This is consistent with the observed stronger dependence of $J$ 
for the SSH model shown in Fig. 2(a). 
The absence of an explicit spin-phonon coupling in leading order
in the HH model and its presence in the SSH model also have 
interesting implications for one dimensional systems.
It has been shown \cite{ogatashiba} that for $U \rightarrow \infty$
the ground state wave function of the 1D Hubbard model may be factorized 
into a product of charge and spin parts. This will still hold including 
electron-phonon coupling of the HH type, but will cease to be valid for
the SSH model, since it is not possible to transform the spin
degrees of freedom into a ``squeezed'' Heisenberg chain in this case.

In summary, we have shown that for two different models of the 
electron-phonon coupling the presence of this interaction
always tends to enhance the effective exchange interaction 
within a single-band model.  In the HH model, if the Coulomb repulsion is
sufficiently large so that the system is still reasonably far
away from the transition to the local bipolaron state
the size of the enhancement is small, so that estimates 
for model parameters are not strongly affected.
The SSH model on the other hand, exhibits rather strong spin-phonon
coupling. 
In particular from Fig. 2(a) one can see that for $\omega_0 = 0.2t$ and
$g = 0.2t$ which is still {\it less} than the critical value for
the formation of a small single polaron (in one dimension)
\cite{ourselves} the enhancement can be nearly a factor of two.
This difference between the HH and the SSH models is a remarkable
feature of the {\it single-band} models only including one
orbital per site, that we considered here.
Derivations of  effective single-band models from multi-band
Hubbard models with general $e$-$ph$ couplings  \cite{SF}
aiming to describe the High $T_c$ superconducting cuprates
show that the effective single-band model should involve
predominantly a Holstein type of coupling. Nevertheless,
our results show that, in case different models or alternative
treatments were to lead to substantial SSH type of coupling
in the effective single-band model, magnetic properties should be
strongly affected since a strong spin-phonon coupling would
result.
We notice, however, that such a strong
spin-phonon coupling should appear in the real high $T_c$ systems as an
isotopic dependence of the observed J and as a dependence of some
phonon frequencies on the magnetization when the temperature
is lowered below the Ne\'el temperature in the antiferromagnetic
insulating compounds. At the moment, to our knowledge, these
features have not been observed in the high $T_c$ systems.

\vfill \eject

\centerline{\bf { FIGURE CAPTIONS}}

    Fig. 1: 

      (a) Effective exchange interaction from exact diagonalization
      of two-site cluster vs. electron-phonon coupling
      for the HH model for $U/t = 16$.
      Curves are labelled by the phonon frequency $\omega_0 /t$
      and the dotted portions indicate the region of on-site
      bipolaron formation.

      (b) Effective exchange interaction vs. electron-phonon coupling
      for the HH model for $\omega_0 /t = 1$.
      Solid lines are from exact diagonalization of two sites,
      dashed are the approximation $J_{PT}$ and dotted lines
      are $J_{SW}$.

    Fig. 2: 

      (a) Effective exchange interaction from exact diagonalization
      of two-site cluster vs. electron-phonon coupling
      for the SSH model for $U/t = 16$.
      Curves are labelled by the phonon frequency $\omega_0 /t$
      and the dotted portions indicate the region of on-site
      bipolaron formation.
      
      (b) Effective exchange interaction vs. electron-phonon coupling
      for the SSH model for $\omega_0 /t = 1$.
      Solid lines are from exact diagonalization of two sites,
      dashed are the approximation (\ref{JSSH}).
\vfill \eject


\begin{thebibliography}{999}
\bibitem{CC} P. Calvani, M. Capizzi, S. Lupi, P. Maselli, A. Paolone and P. Roy,
Phys.Rev.B {\bf 53} (1996) 1.
\bibitem{Holstein} T. Holstein, Ann. Phys. (N.Y.) \vol{8}, 325 (1959);
		   \vol{8} (1959) 343.
\bibitem{polaron_review} R. Micnas, J. Ranninger, S. Robaszkiewicz, Rev.Mod.Phys., {\bf{62}} (1990) 113.
\bibitem{SSH} W.P. Su, J.R. Schrieffer, and A.J. Heeger, Phys. Rev. Lett.\vol{42}
              (1979) 1698.
\bibitem{ZS} J. Zhong and H.-B. Sch\"uttler, Phys. Rev. Lett. \vol{69} (1992) 1600.
\bibitem{proc} H.-B. Sch\"uttler, J.Zhong and A.J.Fedro, in {\it Electronic
Properties and Mechanism of High $T_c$ Superconductors} edited by 
T.Oguchi, K.Kadowaki and T.Sasaki, pp. 295-299 (Elsevier Science Publishers, 1992);

H.-B. Sch\"uttler and A.J. Fedro, Phys. Rev. B \vol{38} (1988) 9063;
\bibitem{IF} D. Ihle and H. Fehske, J. Phys. A. Math. Gen.\vol{28} (1995) 275
\bibitem{LF} I.G. Lang and Yu.A. Firsov, Zh. Eksp. Teor. Fiz. \vol{43} (1962)
 1843            [Sov. Phys. JETP \vol{16} (1963) 1301].
\bibitem {RT} J. Ranninger and U. Thibblin, Phys. Rev. B \vol{42} (1990) 2416
\bibitem{AKR} A.S. Alexandrov, V.V. Kabanov, and D.K. Ray, Phys. Rev. B \vol{49}, 
              (1994) 9915; F. Marsiglio, Phys. Lett. A \vol{180} (1993) 280. 
\bibitem{JJWS}  J. H. Jefferson and W. Stephan, 
Physica {\bf C} 235-240, (1994) 2251.
\bibitem{Harris_Lang} A.B. Harris and R.V. Lange, Phys. Rev. \vol{157}, (1967) 295.
\bibitem{ogatashiba} M. Ogata and H. Shiba, Phys.Rev.B {\bf 41} (1990) 2326.
\bibitem{ourselves} M.Capone, M.Grilli and W.Stephan, preprint
\bibitem{SF} H.-B. Sch\"uttler and A.J. Fedro, Physica {\bf C} 185 (1991) 1673. 
\end{thebibliography}
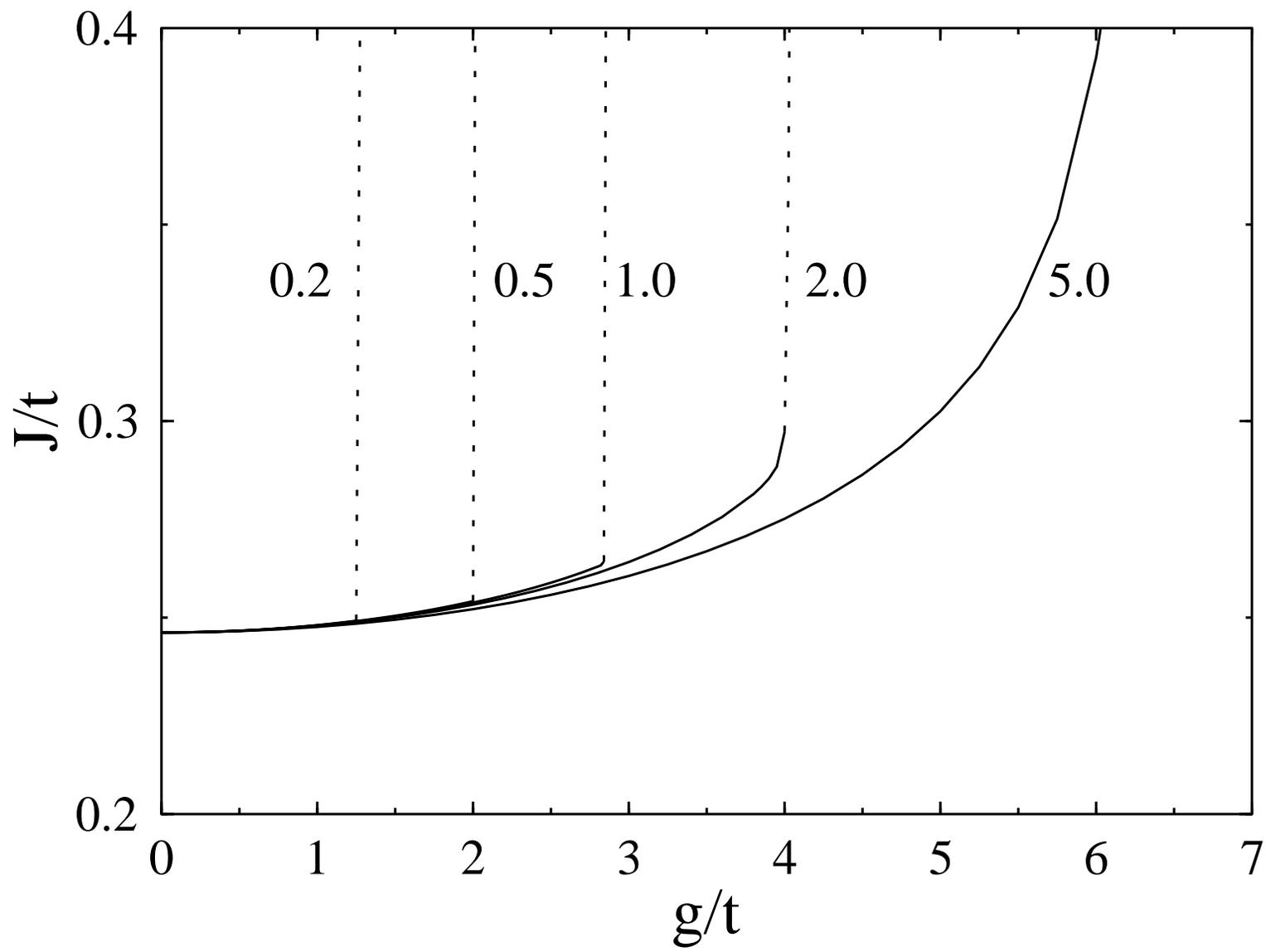
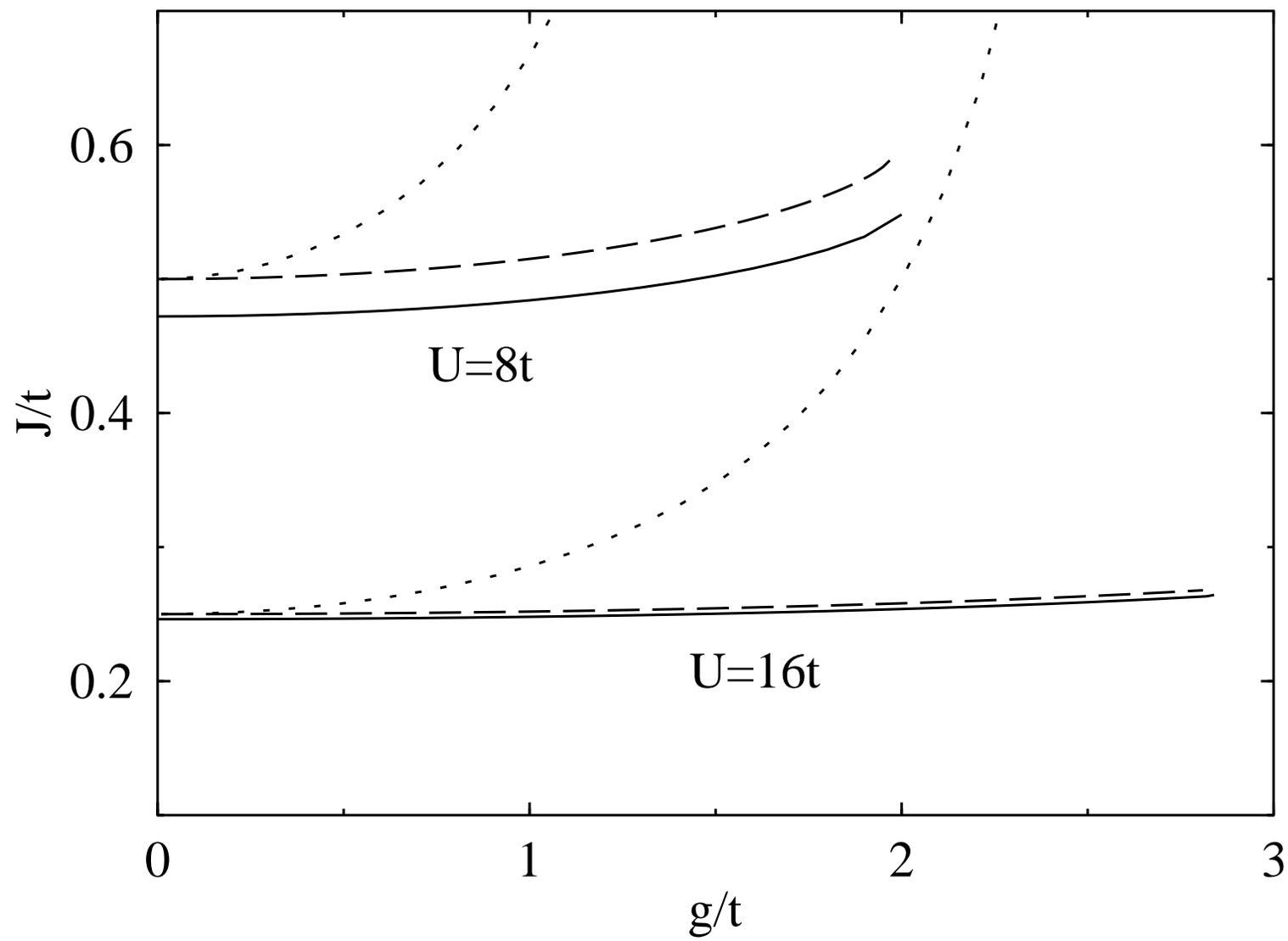
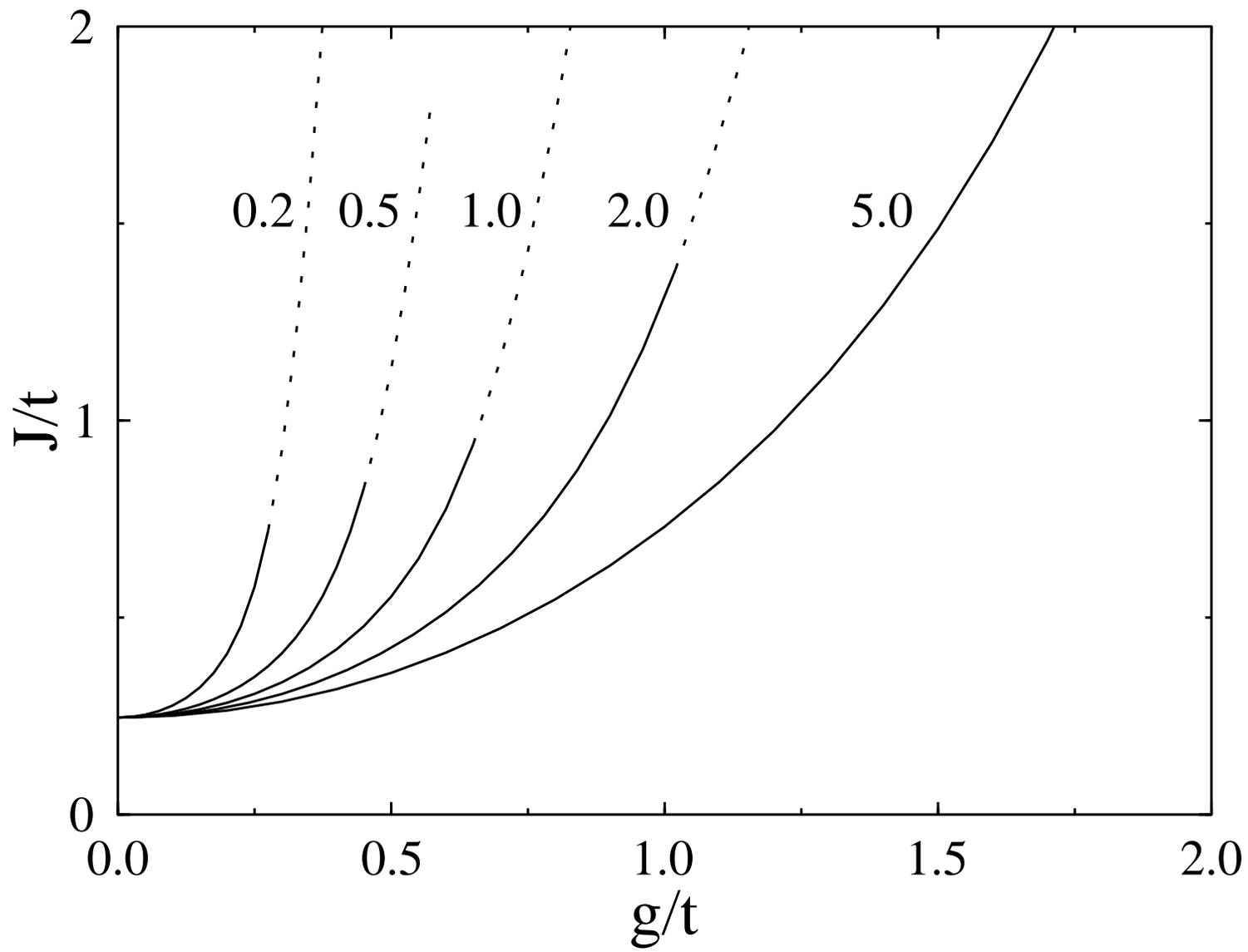
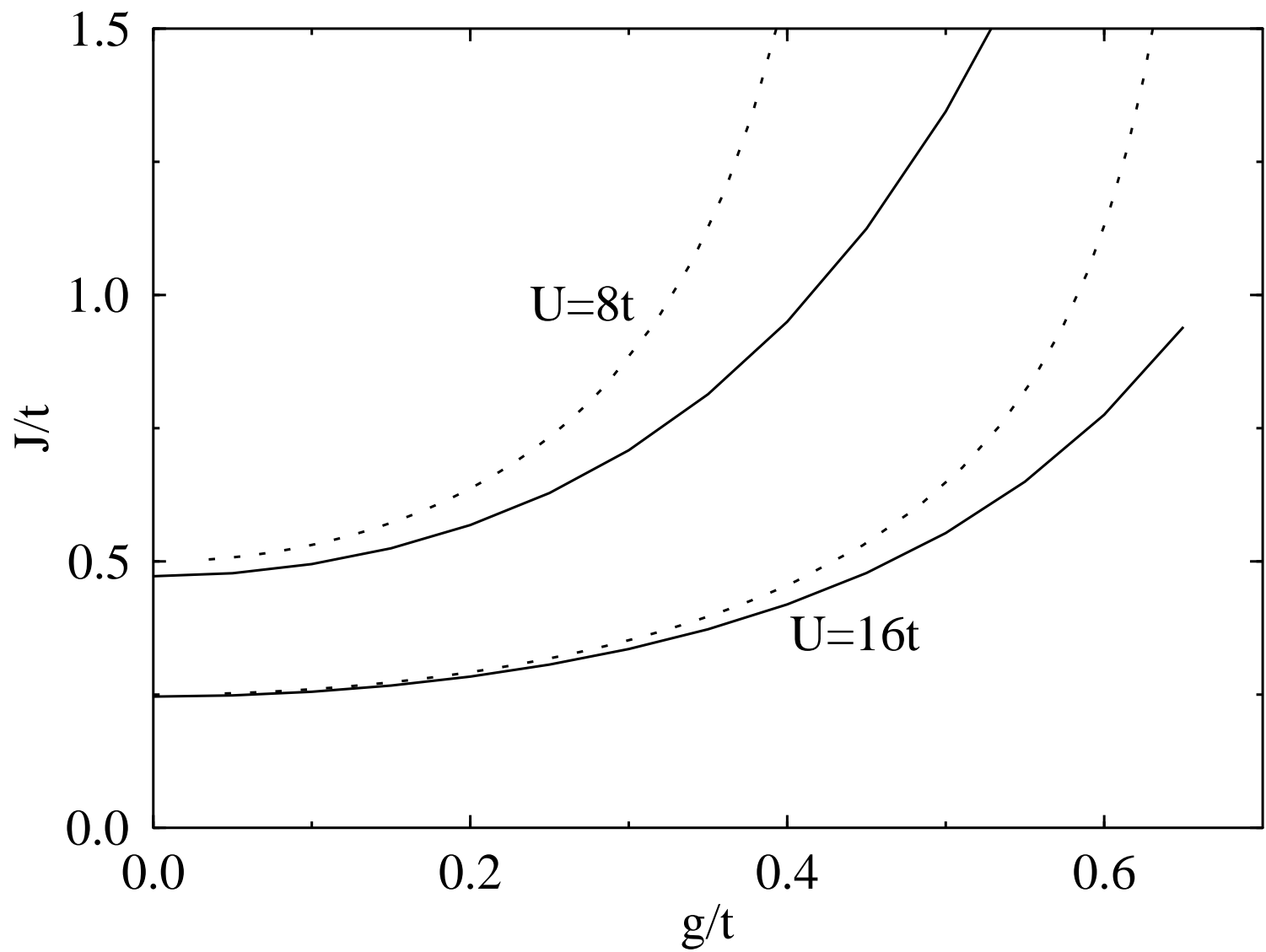
\end{document}